\documentstyle[12pt]{article}

\newcommand{\TlV}{T^{(v)}_{\;\;\; ij}}
\newcommand{\TlM}{T^{(m)}_{\;\;\; ij}}
\newcommand{\be}{\begin{equation}}
\newcommand{\ee}{\end{equation}}

\newcommand{\rf}[1]{(\ref{eq:#1})}

\newcommand{\st}{\stackrel}

\begin{document}
\setlength{\textwidth}{15cm}
\setlength{\textheight}{21cm}
\setlength{\baselineskip}{7mm}

\title{Geometrical Constraints\\on the\\ Cosmological Constant}

\author{M. D. Maia\thanks{E-Mail: MARCOS@FNAlV.FNAL.GOV}\\
and G.S. Silva\\
Universidade de Bras\'{\i}lia\\ Departamento de Matem\'{a}tica\\
70.919 - Bras\'ilia, DF. Brasil\\
and\\
NASA/Fermilab Astrophysics Center\\
Fermi National Accelerator Laboratory\\
Box 500, Batavia, IL 60510-0500}

\maketitle
\begin{abstract}
The cosmological constant problem is examined under the assumption that the
extrinsic curvature of the  space-time contributes to the vacuum. A
compensation mechanism based on a  variable cosmological term is proposed.
Under  a  suitable hypothesis on the behavior of the  extrinsic curvature,
we find that  an initially  large  $\Lambda(t)$ rolls  down rapidly to zero
during the  early stages of the  universe. Using perturbation analysis, it
is  shown that such vacuum behaves essentially as a spin-2 field  which is
independent of the metric.

\end{abstract}

\newpage

\section{Introduction}
Current astrophysical data based on the Friedmann-Robertson-Walker (FRW)
cosmological model estimate an upper bound value for the effective
cosmological constant, so that $\Lambda_{eff}/8\pi\,G $ is very small, to
the
order of $10^{-47} \mbox{Gev}^{4} $. In order to match the contributions
to the vacuum energy density arising from gravity coupled fields, of
the order of $<\rho_{v}>\approx 10^{71} \mbox{Gev}^{4}$, the bare
cosmological constant $\Lambda_{0}$ in Einstein's equations would have
to cancel a very large difference of around 118 decimal places, requiring
an
endless fine tuning in a continuously expanding universe \cite{Wei:1}.

This fundamental problem could in principle be solved by the use of a
simple and elegant generalization of Einstein's equations given by
\begin{equation}
\label{eq:EE1}R_{ij}-\frac{1}{2}Rg_{ij} = 8 \pi G\, t_{ij}
\end{equation}
where
\be
t_{ij} =T^{(v)}_{\;\;\; ij}+T^{(m)}_{\;\;\;ij},
\;\;\;t^{ij}{}_{;j}=T^{(v)\,ij}{}_{;j}+ T^{(m)\,ij}{}_{;j}=0.
\ee
Here $T^{(m)}_{\;\;\; ij}$ denotes the usual energy-momentum tensor of
matter and $T^{(v)}_{\;\;\; ij}$ is the vacuum energy-momentum tensor,
including $\Lambda_{0}$ and all other possible contributions to the vacuum
energy density.  The general form of $T^{(v)}_{\;\;\; ij}$ is not known but
in particular it includes the  so called  variable $\Lambda$ models, where
$T^{(v)}_{\;\;\; ij}=\frac{1}{8\pi G}\Lambda(t)g_{ij}$, for some smooth
scalar function of time $\Lambda(t)$. In this case the general conservation
law in (\ref{eq:EE1}) becomes simply \cite{Oz:1}
\begin{equation}
\label{eq:CONS}\dot{\Lambda}=-g_{0i}\,T^{(m)ij}{}_{;j}.
\end{equation}
To complete our set of equations, we add a state equation generically
expressed as

\begin{equation}
\label{eq:STAT}p =(\gamma -1)\rho,
\end{equation}
In periods of time when matter
becomes almost absolutely conserved we obtain $T^{(m)ij\,}{}_{;j}\approx 0$
and $\Lambda (t)\approx\mbox{constant}$, which could in principle be
adjusted to match the observed cosmological constant.

In spite of many efforts, it appears to be difficult if not impossible to
find a scalar field $\Lambda (t)$ satisfying the appropriate Klein-Gordon's
equation in curved space-time with the total energy momentum tensor as a
source \cite{Wei:1}. This suggets that the vacuum could be a field of
different nature \cite{Dolgov:1}. For  example, from (\ref{eq:EE1}) we see
that $\Lambda (t)$ should behave just like the scalar curvature $R$
\cite{Bar:1}. Nonetheless, we cannot use any linear combination of $R$ as
$\Lambda(t)$, simply because $R$  is already included in left hand side of
(\ref{eq:EE1}).

In another  example $\Lambda (t)$ is taken to be  a function of the
space-time metric, so that its dynamics follows from the generalized
gravitational field equations (\ref{eq:EE1}). Along this line some
interesting models have been proposed \cite{Wu:1},\cite{Waga:1}.
Nonetheless, each of these models use its own ad hoc ansatz for
$\Lambda(t)$ and we still lack a proper explanation for this function, or
more generally for the vacuum tensor $\TlV$.

Assuming that the vacuum is of geometrical nature, then it  should reflect
some geometrical property of the space-time which is not included in the
left hand side of \rf{EE1}. For example, $\Lambda (t)$ could be  made of  a
higher order contraction of curvature tensors, or even from some other
manifestation of the space-time geometry, such as the extrinsic curvature.
In this case, the observed small value of $\Lambda_{eff}/8\pi G$ would be a
symptom of the weak bending of the universe with respect to its tangent
plane.

Based on the fact that the scalar curvature of the space-time may be
decomposed into extrinsic components, we explore the possibility that the
vacuum energy momentum tensor is related to the extrinsic curvature. In the
next section we express the vacuum tensor $T_{\;\;\;ij}^{(v)}$ in terms of
the extrinsic curvature of the space-time. Section 3 describes several
extensions of the FRW model adapted to the more general equations
(\ref{eq:EE1}). Finally, using a perturbative analysis we show in section
four that the extrinsic curvature behaves as a spin-2 field, generally
independent of the metric.

\section{Embedded Cosmologies}

If a given manifold is bent
but not stretched, its intrinsic curvature does not change but its
extrinsic curvature, which measures the deviation from the tangent plane,
would change. Thus, two distinct cosmological models with the same
intrinsic curvatures could exibit different bendings. We may ask if that
difference would have any intrinsically observable effect. We will see that
for the same matter distribution, the difference between the two extrinsic
curvatures corresponds to different vacuum configurations. This means that
the vacuum tensor can be written explicitly in terms of that  extrinsic
curvature (sometimes refered to as the second fundamental form).

Consequently, the observed $\Lambda(t)$ can be taken as an experimental
evidence of the  bending of the universe. Reciprocally,  by specifying the
dynamics of the extrinsic curvature we would be able to describe a number
of interesting scenarios where $\lambda(t)$ rolls down to zero, remaining
in that condition untill today..

The existence of this extrinsic curvature is a consequence of the well know
mathematical property that any manifold, including space-times, can always
be regarded as a subspace of a 5-dimensional space. Therefore, such
subspace structure is not only compatible with, but also complementary to
the exclusively metric description of space-time.

To understand the nature of the extrinsic curvature, we introduce a set of
coordinates $X^\mu ,$ in a non flat 5-dimensional manifold ${\cal M}_5$,
such that\footnote{Most known embedding spaces  ${\cal M}_{D}$ are flat
with $D\le 10$. Here for generality we assume  a curved  space with $D=5$,
so that a large class of comological models can be  accomodated. Our
conventions are: Lower case Latin indices run from 0 to 3. All Greek
indices run from 1 to 5. $R_{ij}= g^{mn}R_{mijn}$. The space-time signature
is $ (+\,+\,+\,-)$.   }
\begin{eqnarray}
g_{ij}  =  X^\mu_{,i}X^\nu_{,j}{\cal G}_{\mu\nu},\; \;
N^\mu X^\nu_{,i}{\cal G}_{\mu\nu}   =  0, \; \;
\epsilon=\pm 1  =  N^{\mu} N^{\nu} {\cal G}_{\mu\nu}.
\label{eq:E}
\end{eqnarray}
Here $N^\nu $ are the components of a vector field $N$ orthogonal to the
space-time and ${\cal G}_{\mu \nu }$ are the components of the metric of $
{\cal M}_5$. The necessary and sufficient conditions for the existence of
a
solution $X^\mu $ of (\ref{eq:E}), is that there is a tensor $b_{ij}$
satisfying the Gauss-Codazzi equations:
\begin{eqnarray}
R_{ijkl} &=& 2\epsilon b_{i[k}b_{l]j}+{\cal R}_{ijkl}, \label{eq:G}\\
\nabla_{[k}b_{i]j}& = &{\cal R}_{ijk5} \label{eq:C}
\end{eqnarray}
where ${\cal R}_{ijkl}$ and ${\cal R}_{ijk5}$ represent projections of the
Riemann tensor of ${\cal M}_5$ along the directions $X_{,i}^\mu $ and
$N^\mu$ respectivelly. The covariant derivative $\nabla _i$ is calculated
with respect to $g_{ij}$. In terms of $X^\mu $, the tensor $b_{ij}$ may be
expressed as
\begin{equation}
b_{ij}=-X_{,i}^\mu N^{\nu}_{,j} {\cal G}_{\mu \nu } \label{eq:bij}.
\end{equation}
Contraction of (\ref{eq:G}) gives an  important  equation:
\begin{equation}
\label{eq:EE2}R_{ij}-\frac 12Rg_{ij}=t_{ij}(b)+\frac 45({\cal
R}_{ij}-\frac
25{\cal R}g_{ij}),
\end{equation}
where ${\cal R}_{ij}$ and ${\cal R}$ are the Ricci tensor and scalar
curvatures of ${\cal M}_5$ respectively, and
\begin{equation}
\label{eq:TAU}t_{ij}(b)=\epsilon(b_{im}b_j^{\;\;m}-hb_{ij}-
\frac 12(\kappa^2-h^2)g_{ij})
\end{equation}
where
$$\kappa ^2=\,b_{mi}b^{mi},\;\;\;h^2=(g^{ij}b_{ij})^2,$$
are the extrinsic scalar and the mean curvatures respectively. Comparing
with (\ref{eq:EE1}) we obtain

\be
T_{\;\;\;ij}^{(v)}=\frac{t_{ij}(b)}{8{\pi}G}+\frac{1}{8\pi
G}\frac{4}{5}({\cal R}_{ij}-\frac{2}{5}{\cal R}g_{ij})-T_{ij}^{(m)}
\ee
and the cosmological function in terms of $b_{ij}$ is defined by
\be
\Lambda(t)\stackrel{def}{=}2\pi G\; tr(T_{\;\;\;ij}^{(v)})=
\frac{\epsilon}{4}(h^2-\kappa^2)-\frac{4}{25} {\cal R}-2\pi G\;
T^{(m)i}{}_i.\label{eq:LAM}
\ee
{}From the contracted Bianchi identity, it follows that the generalized
conservation rule in (\ref{eq:EE1}) is equivalent to the identity
\begin{equation}
\label{eq:STATS}t^{ij}(b)_{\, ;j}=0.
\end{equation}
The dynamics of the field $b_{ij}$ is constrained by the
Codazzi equation (\ref{eq:C}) which has no corresponding in the pure
Riemannian geometry. Equations (\ref{eq:EE2}) say that the space-time
curvature results from the blend of matter and vacuum energy densities,
represented by $t_{ij}(b)$.

\section{FRW Example}
As an illustration, consider the Friedmann-Robertson-Walker (FRW)
parametrization, with coordinates $(x^1, x^2, x^3, x^4) = ( r,
\theta, \phi , t )$:
\begin{equation}
\label{eq:FRW}ds^{2}=-dt^{2}+a^{2}(t)[dr^{2}+f^{2}(r)(d\theta
^{2}+ \mbox{sin}^{2}\theta d\phi ^{2})]
\end{equation}
with $f(r) = r, \mbox{sin}\, r, \mbox{sinh}\, r $ corresponding to $k =0,
+1, -1$ (spatially flat, closed, open respectively). We use the
energy-momentum tensor:
$$
T^{(m)\,ij}= pg^{ij}+(p+\rho) U^{i}U^{j},\;\;
g_{ij}U^{i}U^{j}= -1.
$$
satisfying the more general conservation law in equations in
(\ref{eq:EE1}):
\[
\dot{\Lambda}(t)=8\pi\,G(\dot{\rho}+3H\gamma\rho).
\]
This cosmological model can be embedded in a 5-dimensional flat space with
metric signature $(++++-)$ so that $g_{55}=\epsilon =+1$ \cite{Ros:1}. It
follows that equations (~\ref{eq:G},\ref{eq:C}), reduce to
\begin{eqnarray}
R_{ijkl} & =& 2b_{i[k}b_{l]j},\label{eq:G1}\\
b_{k[i;j]} & = &  0.   \label{eq:C1}
\end{eqnarray}
and the contracted form of \rf{G1} is
\be
R_{ij}-\frac{1}{2}Rg_{ij} =t_{ij}(b) \label{eq:EG}.
\ee
Instead of  calculating  $b_{ij}$  from \rf{bij}, we may  solve
(\ref{eq:C1}) directly obtaining
\begin{eqnarray}
b_{11} = b(t), \; b_{22} = f^{2}(r)b(t), \; b_{33}
=(f(r)\,\mbox{sin}\theta)^{2}b(t),\;
b_{00}=-\frac{1}{\dot{a}}\frac{d}{dt}\big{(}\frac{b}{a}\big{)}\label{eq:BIJ}
,
\end{eqnarray}
where $b(t)$ is an arbitrary function of $t$ and all other components are
zero. Replacing these values in (\ref{eq:EG}), we obtain
\begin{eqnarray}
\frac{\ddot{a}}{a}=\frac{b}{a^{2}\dot{a}}\frac{d}{dt}(\frac{b}{a}), \\
a\ddot{a}+2\dot{a}^{2} +2k=\frac{b}{a^{2}\dot{a}}\frac{d}{dt}(ab).
\end{eqnarray}
While the left hand sides of these equations are the familiar FRW
expressions, the right hand sides, built with $b_{ij}$ are new. Eliminating
$\ddot a$ among those equations, we obtain a single relation between $b$
and $a$:
\begin{equation}
\label{eq:fri}\dot a^2+k=\frac{b^2}{a^2}.
\end{equation}
Denoting the Hubble parameter $H=\frac{\dot a}a $ and the relative change
of the extrinsic curvature by $B=\frac{\dot b}b $, we may write
\be
\label{eq:BB}b_{ab}=\frac{b}{a^2}g_{ab},\;\;
b_{00}=\frac{b}{a^{2}}(\frac{B}{H}-1)g_{00}, \;\;a,b=1..3
\ee
The extrinsic scalar and mean curvatures are respectively
\begin{equation}
\kappa^{2}=\frac{b^2}{a^4}\left( \frac{B^2}{H^2}-2\frac BH+4\right)
\;\;\;\;\;h=\frac b{a^2}\left(\frac BH+2\right)\label{eq:hk}
\end{equation}
and the vacuum energy-momentum tensor in terms of $b(t)$ is given by
\begin{eqnarray}
T^{(v)}{}_{ab} & = & \frac{1}{8\pi G}\frac{b^{2}}{a^4}\left(
\frac{2B}{H}-1\right ) g_{ab} - p g_{ab},\nonumber \\
T^{(v)}{}_{00}& = & -\frac{1}{8\pi G}\frac{3b^{2}}{a^{4}} -\rho.\nonumber
\end{eqnarray}
The cosmological function \rf{LAM} expressed in terms of the extrinsic
geometry is
\begin{equation}
\Lambda(t)=\frac{6b^2}{4a^4}\frac{B}{H}-2\pi G\left(3p -\rho\right) .
\end{equation}
Let us examine the meaning of the extrinsic curvature in some particular
situations. Defining a new function $\omega (t)$ by $b=e^\omega a^2$, it
follows that
$$
B=2H+\dot \omega
$$
and (\ref{eq:fri}) becomes $\dot a^2+k=e^{2\omega }a^2$. Comparing with
the
extended Friedmann equation (with $\Lambda (t)$)
\begin{equation}
\label{eq:exfri}\dot a+k=\frac{8\pi G}3(\rho -\frac{\Lambda (t)}{8\pi
G})a^2,
\end{equation}
we obtain $e^{2\omega }=\frac{8\pi G}3\rho _T$, where $\rho _T=\rho
-\Lambda (t)/8{\pi }G$ is the total energy density. The derivative of this
equation gives
$$
\frac 12\frac{\dot{\rho}_{T} }{\rho _T} =\dot \omega = B-2H.
$$
This can be integrated producing the general expression for the matter
energy density in terms of $B$, $H$ and $\Lambda(t)$:
\be
\rho_{T}=\rho_{0T}e^{\int{2(B-2H)dt}} \label{eq:RHO}
\ee
where $\rho_{0T}$ is  an integration constant.
In particular, when $\omega =\omega _0=\mbox{constant}$, that is when the
extrinsic curvature is such that $B=2H$, then  $\rho $ and  $\Lambda(t)$
become constants. It also follows from (\ref{eq:BB}) that in this case
$b_{ij}$ is conformally related to the metric:
$$
b_{ij}=e^{\omega _0}g_{ij}
$$
Replacing $b_{ij}$ in (\ref{eq:G1}), we obtain
$$
R_{ijkl}=2e^{2\omega _0}g_{i[k}g_{l]j}
$$
showing that the space-time has constant curvature, which means in our
example a  deSitter universe.
Let us suppose now that $\omega $ starts fluctuating. For example,  we may
have a membrane model of universe characterized by the condition $h=0$
\cite{Mai:1}. From (\ref{eq:hk}), this implies that $B=-2H$, corresponding
to $\dot\omega <0$. Integrating this relation we obtain $b=\frac{b_0}{a^2}$
and replacing in \rf{RHO}, we obtain a total density decaying as $\frac
1{a^8}$, which means  a universe filled with some kind of hyper-stiff
combination of  vacuum and  matter \cite{whe:1}, decaying to zero at a
later  stage. From  \rf{LAM} we see also that the sign of $\Lambda (t)$ is
governed by the mean curvature of the space-time and the signature of
${\cal M}_{D}$. In a membrane universe  with $h=0$ and $\epsilon=1$, we
would have $\Lambda (t)<0$ \cite{Mai:1}. On the other hand, when $\dot
\omega >0$ or, equivalently when $B>2H$, we have a universe  where
$\Lambda(t)$ and $\rho(t)$ would increase  as positive powers of $a(t)$,
leading to somewhat strange universes
without the primordial singularity.

So far  we have used only  a general theorem on submanifolds to  cosmology,
without any further assumptions. Let us now consider a condition on the
extrinsic curvature expressed as
\begin{equation}
\label{eq:HIP1}B=\alpha H
\end{equation}
where $b_{0}$ is an integration constant  and where $\alpha $ is some yet
unknown parameter. Since $B$ is a measure of the relative variation of the
normal $N$, and $H$ is always decreasing with
time, (\ref{eq:HIP1}) has the reasonable interpretation that the relative
variation of the normal vector $N$ decreases in proportion to the relative
expansion of the universe. From  \rf{HIP1} we have that $b=b_{0}a^{\alpha}$
so that $\Lambda(t)$ can be written as
\begin{equation}
\label{eq:lambda}\Lambda (t)={\Lambda}_{0}{\alpha}a^{2(\alpha-2)}- 2\pi G
(3p-\rho)
\end{equation}
where ${\Lambda}_{0}$ is a constant.
The total density becomes
$$
\rho_{T}=\rho_{T}a^{2(\alpha-2)}.
$$
When $\alpha =2$, we have  a  constant  curvature space with a constant
density. We  may fix this constant  asuming that it corresponds to the
total critical density: $\rho =\rho_{Tcr}$. In this case, the  relative
total density of vacuum and matter is
\begin{equation}
\Omega_{T}=\frac{\rho_{T}}{\rho_{Tcr}} = a^{2(\alpha-2)}
\end{equation}
As we see,  depending on the value of  $\alpha$  we  obtain different
scenarios,  not necessarily corresponding to   realistic models.
The following table shows the decay $h$, $\Lambda$ and $\Omega_{T}$
corresponding to some values of $\alpha$, where we have set
${\Lambda}_{0}=1$ and used $3p=\rho$:

\vspace{0.5cm}
\begin{tabular}{|c|c|c||c|c|c||c|c|}
\hline
        & mem-  & stiff vac & radia-. & vac+matt.  & deSitter  & strange &
stranger  \\
        & branes &+ matter &tion & dust & inflation & worlds & worlds\\
\hline
 $\alpha =$ & -2 &-1 & 0 & 1& 2 &3& 4 \\
\hline
$h\leadsto$  & $0$ & $1/{a^{3}}$ & $ 2/a^{2}$ & $ 3/{a}$ & $4$ & $5a$ &
$6a^{2}$\\
\hline
$\Lambda\leadsto$ & $-2/a^{8}$ &
$-1/a^{6}$ & $0$ & $1/a^{2}$& $2$ & $3a^{2}$ & $4a^{4}$ \\
\hline
$\Omega_{T}\leadsto $ & $1/a^{8}$ & $1/a^{6}$ & $1/a^{4}$ & $1/a^{2}$ & $1$
& $a^{2}$ & $a^{4}$\\
\hline
\end{tabular}
\begin{center}
{\small table1: some values of $\alpha $ and corresponding scenarios}
\end{center}

\vspace{0.3mm}
We can now replace  $b(t)$ in \rf{fri} to obtain the  expansion  equation
($\alpha =$ constant)
\be
\dot{a}^{2} + \kappa = b_{0}^{2} a^{2(\alpha-1)}
\ee
It is quite conceivable that the proportion between $B$ and $H$ vary in
different periods of the history of the universe. As table 1 suggests,
$\alpha $ could be a function of time. One possible sequence would be that
$\alpha \st{>}{\approx}2$ at the beginning of the universe, with a density
incresing at the rate of $a^{2}$, followed by a deSitter inflation  with
$\alpha =2$, $\Lambda =$ constant  and  $\Omega_{T}=1$. Next we would have
a post inflationary period  with $\alpha \st{<}{\approx}1$ when $\Lambda
\leadsto 1/a^{2}$. The other remaining values of $\alpha$ may not
correspond to physically interesting models.

The following graphs\footnote{The $a$  scale has been roughly divided in 3
parts only: the  early,  middle and late universe.  For plotting
compatibility we have eliminated the singularity at $a=0\;$ \cite{Wolf:1}.}
display $\Lambda(t)$ and  $\Omega (t)$ as a functions of the expansion
factor  $a(t)$ and of a possible  continuous parameter $\alpha $.
As we see in Fig.1, the  strip between $\alpha \approx 0$ and the inflation
line $\alpha \approx 2$ corresponds to the most common scenarios. In this
region, $\Lambda(t)$ starts  as a positive value  but fall rapidly to zero
during the early stages of the universe, remaining in that condition
untill today. The values  $\alpha>2$  represent   a region where
$\Lambda(t)$  has a polynomial increases at late universe. On the other
hand, the regions located  at $\alpha <0$ have an opposite  result, where
$\Lambda(t)$ starts with negative values but rapidly  increases to zero at
the early stages of the universe.

\vspace{8cm}
\begin{center}
{\small Fig.1: $\Lambda$ as continuous functions of $\alpha$ and $a$}
\end{center}
Figure 2 shows the behavior of the relative total density $\Omega_{T}$ as a
function of  $a(t)$ and $\alpha$. Again,  not all values of  $\alpha$
corresponds to realistic models. From \rf{HIP1} we see that for the
deSitter inflation line at $\alpha =2$ we have $\Omega_{T} =1$,  suggesting
that at the early universe $\alpha$ could have been slightly larger than 2.
If we assume that  the observed density is  the  result of the
contribution of  the vacuum and matter, that is, $\rho_{T}$,  then we may
take that $0.1 \le \Omega_{T} \le 2$. Consequently, today's universe would
be somewhere in the region defined by  $1 <\alpha <2$, which in accordance
with Fig.1 predicts  $\Lambda (t) \approx 0$ at  late universe. The non
integer values of $\alpha$ in this domain seem to fit  in a large classe
of acceptable models.
\newpage
{}.

\vspace{8cm}
\begin{center}
{\small Fig.2:  $\Omega_{T}$ as continuous functions of $\alpha$ and $a$}
\end{center}

\section{Spin-2 Vacuum}
The  hypothesis \rf{HIP1}
should be replaced by  a dynamical  equation  for the extrinsic curvature.

The obvious candidate for the Lagrangia would be constructed with the
second invariant of the extrinsic curvature $K=\epsilon(\kappa ^2 -h^2)$,
so that $K\sqrt{-g}$ is the extrinsic version of the Einstein-Hilbert
Lagrangian. As it happens, the Euler-Lagrange equations derived from this
expression with respect to $g_{ij}$ does not reproduce $t_{ij}(b)$. This
can be explained by the fact that $K$ is homogeneous of degree 2 in
$g_{ij}$, while Einstein-Hilbert's Lagrangian is homogeneous of degree -1,
leading to an inconsistency. To obtain the correct Lagrangian we notice
that

\be
\frac{\partial}{\partial g^{ij}}
(\sqrt{K}\sqrt{-g})=\frac{1}{\sqrt{K}}t_{ij}(b)\sqrt{-g}
\ee
Therefore a Lagrangian which does not depend on derivatives of $g_{ij}$
would be given by the classical path integral over all embedded geometries:
\begin{equation}
\label{eq:L1}{\cal L}_{1}=\int {\sqrt{K}\frac \partial {\partial
g^{ij}}(\sqrt{K}\sqrt{-g})\,dg_{ij}},
\end{equation}
We can easily see that the Euler-Lagrange equations for $R\sqrt{-g}-{\cal
L}_{1}$ with respect to $g_{ij}$ reproduce equation \rf{EG}. On the other
hand, the Euler-Lagrange equations with respect to $b_{ij}$ gives only an
algebraic, expression (that is, not involving derivatives) on  this field,
so that ${\cal L}_{1}$ is not telling the whole story.  To write the
complete Lagrangian we need a deeper insight on the nature of the field
$b_{ij}$, which can be obtained  from a perturbative analysis addapted to
the embedded cosmological models.

A space-time perturbation may be intuitively conceived as a local growth
or
deformation of a given background geometry. Mathematically
speaking, this deformation may be described by a shift of the background
along some transverse (that is, not tangent) vector field $\zeta $,
producing a one parameter family of manifolds. The perturbed or physical
space-time is a
manifold with the same differentiable structure as the background, whose
points are identified with all points along the integral curve of $\zeta$.
Therefore, the metric of the perturbed manifold depends on the family
parameter and it may be calculated by the Lie transport of the background
metric along the integral curve \cite{Ger:1}. Thus, different
perturbations are generated by different choices of transverse vectors
$\zeta$.
Since this is a transverse vector, its tangent component induce a
coordinate
transformations, producing a ``coordinate gauge'' condition on the
perturbation. Of course, this is undesirable and should be filtered out.
Only the perturbations which are independent of coordinate gauges
are physically meaningful \cite{Wal:1} and correspond to  a density
perturbation \cite{Bardeen:1},\cite{Branden:1}. Although this
perturbation is described as a classical process, it eventually started
from the backreactions of the fluctuations of quantum fields interacting
with the classical gravitational field of the background.
Here we  merelly apply the above definition to the case of  a space-time
perturbation given by the  tensor $b_{ij}$.

A family of embedded submanifolds of the flat space ${\cal M}_5$, in the
neighborhood of that space-time, may be described by the Cartesian
coordinates $$
Z^\mu =X^\mu +sN^\mu,
$$
where $s$ is a parameter such that $s=0$ correspond to the original
background space-time. In this case, a transverse vector can be written as
$$
\zeta^{\mu} =c^iX^{\mu}{}_{,i}+cN^{\mu},\;\;\;c\neq 0.
$$

If $Q$ is a geometrical object defined in the background, then the
corresponding change of $Q$ produce by the above perturbation is given by
the Lie derivative $\pounds_{\zeta}Q$. For a small value of the parameter
$s$, the linear or first order perturbation is
$$
\stackrel{(1)}{Q}=Q+s\pounds _\zeta Q.
$$
Consider two distinct linear perturbations of the same object $Q$
generated
by two transverse vectors $\zeta$ and $\zeta ^{\prime }$. Then,
$$
\stackrel{(1)}{Q^{\prime }}-\stackrel{(1)}{Q}=\pounds _\xi Q-\pounds _\eta
Q,
$$
where $\xi =(s^{\prime }c^{\prime }{}^i-sc^i)X_{,i}^\mu $ is a tangent
vector and $\eta =(s^{\prime }c^{\prime }-sc)N^\mu $ is a normal vector.
Suppose $\eta $ could be set to zero, without any further
conditions imposed on $\xi $, which remains arbitrary\footnote{This would
be the case if the signature of  $N$ is taken to be zero as in
\cite{Ger:1}, where there is  no measure for the length $c$. }. In this
case, the two perturbations would be equal whenever $\pounds _\xi
Q=0$ for any $\xi $. This implies that we could eliminate the perturbation
by an infinitesimal coordinate transformation. Such condition can
be satisfied only by a very special class of geometric objects. The metric
could never be one such object unless all vector fields of the space-time
are Killing vector fields. One possible solution to the above  coordinate
gauge problem is to take the normal vector $N$ as the transverse vector
(with a non zero signature). In this way, no coordinate transformation can
interfere with the perturbation. Next, to construct the perturbation of an
object, we first determine the perturbation of a tetrad field
$\{h^{i}_{j}\}$ and then contract the object with the perturbed tetrad
\cite{Chandra:1}.
The linear perturbation of $\{h^{i}_{j} \}$ generated by $N$ is
$$
\stackrel{(1)}{h^{i}_{j}}\,=\, h^{i}_{j}+s\pounds _N h^{i}_{j}
$$
and the physical components of  the metric in the perturbed tetrad is given
by
\[
\st{(2)}{g_{ij}}= \st{(1)}{h^{m}_{i}}\st{(1)}{ h^{n}_{j}}g_{mn}.
\]
Since our perturbed space-times have a metric induced by ${\cal G}_{\mu \nu
}$ (or $\eta _{\mu \nu }$ in the flat case), we may   replace the tetrad
$\{h^{i}_{j}\}$  by a vielbein $\{l_i^\mu \}$ and the perturbed metric will
be given by
$$
\stackrel{(2)}{g_{ij}}\,=\,\stackrel{(1)}{l_i^\mu }\;\stackrel{(1)}{l_j^\nu
}
{\cal G}_{\mu \nu }.
$$
Let us  consider the unperturbed vielbein to be $l_i^\mu = X_{\;,i}^\mu$
and,
just for the sake of intuition, that the normal $N$ points towards the most
convex side of the space-time. Then, the perturbed vielbein along this
normal is
$$
\stackrel{(1)}{l_i^\mu }=X_{\;,i}^\mu -s\pounds _NX_{\;,i}^\mu
=X_{\;,i}^\mu
+sN_{\;,i}^\mu
$$
so that the  second order geometric  perturbation of the metric is
\begin{equation}
\stackrel{(2)}{g}_{ij}(x,s)=
\stackrel{(1)}{l_{i}^{\mu}}\,\stackrel{(1)}{l_{j}^{\nu}}
\eta _{\mu \nu }=X_{\;,i}^\mu X_{\;,j}^\nu \eta _{\mu \nu }+2sN_{\;,i}^\mu
X_{\;,j}^\nu \eta _{\mu \nu }+s^2N_{\;,i}^\mu N_{\;,j}^\nu \eta _{\mu \nu
},
\end{equation}
or, using \rf{bij},
\begin{equation}
\label{eq:til}\stackrel{(2)}{g}_{ij}=g_{ij}-2sb_{ij}+s^2g^{mn}b_{im}b_{jn}=g
^{mn}(g_{im}-sb_{im})(g_{jn}-sb_{jn}).
\end{equation}
An approximate field equation for $b_{ij}$ can be obtained from the linear
perturbation of the metric (assuming $s^2<<s$ in  \rf{til}):
$$
\stackrel{(1)}{g_{ij}}\approx g_{ij}-2s\,b_{ij}
$$
Replacing this in \rf{EE1},  and applying the usual deDonder gauge
$(b^{i}_{j}-h/2\, \delta^{i}_{j})_{,i}=0$, we obtain

\be
\Box ^ 2\,(b_{ij} -\frac{1}{2}h\, g_{ij})
-R_{i\;\;\;\,j}^{\,\;k\,l}\,(b_{kl}-\frac{1}{2}h\,g_{kl})=8\pi G(\TlV
+\TlM) \label{eq:WAVE},
\ee
so that the extrinsic curvature is a spin-2 field over the
background with the total energy-momentum tensor as  source. Now, if a
spin-2 field  has  the  total energy momentum tensor  as  a source, it must
necessarily be derived from a  Einstein-Hilbert type Lagrangian ${\cal
L}_{2}=R(b_{ij})\sqrt{-det(b_{ij})}$, where $b_{ij}$ takes the place of the
metric in all expressions  and contractions \cite{Gupta:1}. The resulting
Lagrangian  would therefore be
\be
{\cal L}_{b} ={\cal L}_{1}+{\cal L}_{2}=\int {\sqrt{K}\frac \partial
{\partial g^{ij}}(\sqrt{K}\sqrt{-g})\,dg_{ij}}\;+\;
R(b_{ij})\sqrt{\frac{det(b_{ij})}{g}}\sqrt{-g}
\ee
Notice that the dependence on $g_{ij}$ in the second part of this
Lagrangian is only fictitious. We may now set the total Lagrangian as
\[
{\cal L}= R(g)\sqrt{-g}-{\cal L}_{b}
\]
This Lagrangian, tells us  how the universe bends in response to the
vaccum energy  and in particular determines the behavior of  $\Lambda(t)$.
The corresponding field equations with respect to $g_{ij}$  are

\[
R_{ij}(g)-\frac{1}{2}R(g)g_{ij}= t_{ij}(b)=8\pi G(\TlV +\TlM)
\]
while the field equations with respec to $b_{ij}$  are

\[
R_{ij}(b) -\frac{1}{2}R(b)\, b_{ij}= f_{ij}(b)
\]
where $f_{ij}(b)$  is  defined by
\[
\frac{\partial f_{ij}(b)}{\partial
g^{mn}}=\epsilon\left[g_{(im}b_{n)j}-(b_{ij}-h
\,g_{ij})g_{mn}-(g_{ij}b_{mn} -hg_{im}g_{jn})\right]\sqrt{-g}.
\]
This  tensor is not  a  source  term for  $b_{ij}$ and is not zero in
general.
Further constraints  may be imposed on  the  extrinsic  geometry of the
space-time by   assigning specific values for  $f_{ij}$.
\vspace{1cm}\\
In summary, the  vacuum energy was expressed in terms of the extrinsic
curvature of the space-time. In  a first exploratory analysis  we have used
the geometric constraint $B =\alpha H$, obtaining several possible
scenarios where $\Lambda (t)$ decays rapidly to zero and $\Omega$ becomes 1
or  a  value  smaller than 1 in its late stages. Next, a dynamic  condition
on the extrinsic curvature was   defined by a  perturbation of the
space-time generated by  $b_{ij}$. We concluded that this field  behaves as
a  spin-2 field over the background, with the total energy-momentum tensor
as its source. Once this was understood, we  applied a general theorem due
to  S. Gupta to  derive the finite Lagrangian for $b_{ij}$. In a
subsequent paper,  we will apply those results to examine the density
perturbation induced by  the extrinsic curvature in the FRW example.

\vspace{1cm}

Acknowledgements:
\vspace{.5cm}\\
The authors  wish to thank the  warm hospitality received at Fermilab.
This work was supported in part by grants from CAPES, CNPq, DOE and  NASA
grant \# NAGW-2381.


\newpage

\begin{thebibliography}{99}
\bibitem{Wei:1}  S. Weinberg, Rev. Mod. Phys. {\bf \underline{61}, 1,
(1989)}

\bibitem{Oz:1}M. Ozer \& M.O. Taha. Nucl. Phys {\bf \underline{287}, 777,
(1987)}

\bibitem{Dolgov:1} A. D. Dolgov. The Cosmological Constant Problem\\
    in  {\em The Quest for Fundamental Constants in Cosmology.} J. Audazes
\& J Tran Thanvan (eds). Editions Frontiers, Gif-Sur-Ivette, Cedex  France,
p. 227 {\bf (1989)}
\bibitem{Bar:1}  S. M. Barr. Phys. rev {\bf \underline{D36}, 1691, (1987)}

\bibitem{Wu:1}  Wei Chen \& Yong-Shi Wu. Phys. Rev.{\bf \underline{D41},
695
                (1990) }.
\bibitem{Waga:1}  I.  Waga. The Astrophys. Jour. {\bf \underline{414}, 436
(1993)}

\bibitem{Ros:1}  J. Rosen. Rev. Mod. Phys. {\bf \underline{37}, 204,
(1965)}

\bibitem{whe:1} B. K. Harrison, K.S. Thorne, M. Wakano and J. A. Wheeler.
 {em Graviattion Theory and Gravitational Collapse.} U. of Chicago Press,
Chicago {\bf (1964)}
\bibitem{Mai:1}  M.D. Maia \& W.L. Roque Phys. lett. {\bf \underline{A139}.
121, (1989)}
\bibitem{Wolf:1} S. Wolfram. {\em Mathematica }(2nd ed)  Addison Wesley,
N.Y (1991)

\bibitem{Ger:1}R. Geroch. Commun. Math. Phys.
{\bf\underline{13},180,(1969)}

\bibitem{Wal:1} J.M. Stewart \&  M. Walker. Proc. Roy. Soc. London {\bf
\underline{A341}, 49, (1974)}
\bibitem{Bardeen:1} J.M. Bardeen. Phys. Rev. {\bf\underline{D22}, 1882,
(1980)}
\bibitem{Branden:1}  V. F. Mukanov, H.A. Feldman \& R.I. Brendenberg.
Phys. Rep. {\bf\underline{215},203 (1992)}

\bibitem{Chandra:1} S. Chandresaker. {\em The Mathematical Theory of Black
Holes}. Oxford University Press, Oxford {\bf (1992)}

\bibitem{Gupta:1}  S.N. Gupta.  Phys. Rev {\bf \underline{96}, 1683
(1954)}

\end{thebibliography}
\end{document}